\documentclass[12pt]{iopart}

\usepackage{iopams}  
\usepackage[english]{babel}
\usepackage{graphicx}
\usepackage{color}
\usepackage{bbm}

\begin{document}

%
\title{The quantum Rabi model for two qubits}

\author{S.~A. Chilingaryan}
\address{Departamento de F\'{i}sica, Universidade Federal de Minas Gerais, caixa postal 702, 30123-970, Belo Horizonte, MG, Brazil}
\ead{suren@ufmg.br}

\author{B.~M. Rodr\'{\i}guez-Lara}
\address{Instituto Nacional de Astrof\'{i}sica, \'{O}ptica y Electr\'{o}nica \\ Calle Luis Enrique Erro No. 1, Sta. Ma. Tonantzintla, Pue. CP 72840, M\'{e}xico}
\ead{bmlara@inaoep.mx}

%
\begin{abstract}
We study a system composed of two nonidentical qubits coupled to a single mode quantum field.
We calculate the spectra of the system  in the deep-strong-coupling regime via perturbation theory up to second order corrections, and show that it converges to two forced oscillator chains for cases well into that regime.
Our predictions are confirmed by numerical calculation of the spectra using a parity decomposition of the corresponding Hilbert space. 
The numerical results point to two interesting behaviors in the ultra-strong-coupling regime: the rotating wave approximation is valid for some particular cases and there exist crossings in the spectra within each parity subspace.
We also present the normal modes of the system and give an example of the time evolution of the mean photon number, population inversion, von Neuman entropy and Wootters concurrence in the ultra-strong- and deep-strong-coupling regimes. 
\end{abstract}


\maketitle

%
\section{Introduction}

The Dicke model \cite{Dicke1954p99} is the simplest model describing the interaction of an ensemble of identical two-level systems (qubits) with a single mode boson field.
For weak couplings where the rotating wave approximation (RWA) is valid, it is known as the Tavis-Cummings model \cite{Tavis1968p170}, it is exactly solvable  and has successfully used to describe collective phenomena in quantum electrodynamics (QED) \cite{Garraway2011p1137}.
For a single qubit, the Dicke model has recently been known as the quantum Rabi model or just Rabi model. 
For many years, the Rabi model was just a mathematical curiosity as dipole-field systems could not reach in a simple way the coupling regimes where the RWA breaks.
Nevertheless, the system was extensively studied via Bargman representation \cite{Schweber1967p205}, continued fractions \cite{Swain1973p1919, Tur2000p574, Tur2001p899}, semi-classical methods \cite{Graham1984p233}, perturbation theory \cite{Zaheer1988p1628, Phoenix1989p1163}, and coupled cluster \cite{Bishop1996R4657} methods.

Recently, the ability of solid-state-cavity-QED \cite{Englund2007p857}  and circuit-QED systems to attain coupling regimes where the RWA fails has relighted the interest on the single qubit Rabi model.
Circuit-QED systems can reach couplings greater than the field frequency \cite{Devoret2007p767}; e.g. ultra-strong-coupling (USC)  \cite{Ciuti2005p115303,Bourassa2009p032109, Niemczyk2010p772}, where couplings are of the order of tens of the field frequency, $g/\omega_{f} \sim 0.1$, and deep-strong-coupling (DSC) \cite{Casanova2010p263603}, where couplings are comparable or larger than the field frequency, $g/\omega_{f} \gtrsim 1$. 
Thus, the single qubit Rabi model has been revisited using perturbation theory \cite{Casanova2010p263603, Yu2012p015803}, continued fractions \cite{Pan2010p175501, Ziegler2012p452001},  Bargmann representation \cite{Braak2011p100401, Moroz2012p12053139, Maciejewski2012p12101130, Braak2012p12204946, Maciejewski2012p12114639, Wolf2012p053817, Wolf2013p023835}, and coherent states \cite{Chen2012p023822} methods.

Here, we are interested in a experimentally feasible circuit-QED system composed of two nonidentical superconducting qubits coupled to a strip-line resonator. 
In the weak coupling regime and for identical qubits, this system has been of use in quantum information theory for coherent storage and transfer between two phase qubits  \cite{Sillanpaa2007p438} and for resonant two-qubit phase gates \cite{Haack2010p024514}; the validity of the latter breaks for coupling regimes where the full Dicke model has to be taken into account \cite{Wang2012p014031}.
The most general model is described by the Hamiltonian, 
\begin{eqnarray} \label{eq:Hamiltonian}
\hat{H} = \omega_{f} \hat{a}^{\dagger} \hat{a} + \frac{1}{2} \left(\omega_{1} \hat{\sigma}_{z}^{(1)} + \omega_{2} \hat{\sigma}_{z}^{(2)}\right) +  \left(\hat{a} + \hat{a}^{\dagger} \right) \left( g_{1} \hat{\sigma}_{x}^{(1)} + g_{2} \hat{\sigma}_{x}^{(2)}\right),
\end{eqnarray}
 where the two nonidentical qubits are depicted by the Pauli operators $\hat{\sigma}_{j}^{(k)}$ with $j=x,y,z$ and $k=1,2$ and the transition frequencies $\omega_{1}$ and $\omega_{2}$.
 The single mode quantum field is described by the creation (annihilation) operators $\hat{a}^{\dagger}$ $(\hat{a})$ and the frequency $\omega_{f}$.
 This model conserves parity, $[\hat{\Pi}, \hat{H}] = 0$, defined as
 \begin{eqnarray}
 \hat{\Pi} = \hat{\sigma}_{z}^{(1)} \hat{\sigma}_{z}^{(2)} (-1)^{\hat{a}^{\dagger} \hat{a}}. \label{eq:Parity}
 \end{eqnarray}
Up to our knowledge, such a system has only been studied in two simplified forms: one for identical qubits with frequencies much smaller than the oscillator frequency \cite{Agarwal2012p043815} and the other for non-identical qubits where the transition frequency of one of them is set to zero \cite{Peng2012p365302}.
Also, particular initial states and parameter values have been used to explore the effect of counter rotating terms on entanglement and discord in closed  \cite{Chen2010p052306} and lossy \cite{Ficek2010p014005,Altintas2012p1791} two-qubit quantum Rabi systems.

In the following, we discuss the exact and approximate eigenvalues of Hamiltonian (\ref{eq:Hamiltonian}) in the weak- and DSC regimes, in that order. 
Curiously, we find that the validity of the RWA can be extended into the USC regime by keeping one coupling parameter value low or using symmetric detunings in the qubits.
We also find that for large values of the coupling parameters, $g_{1}, g_{2} \gg \omega_{1}, \omega_{2}$, the eigenvalues are described by the spectra of two forced oscillators.
We take advantage of parity conservation to numerically calculate the spectra in each of the parity subspaces.
An important feature of the full two-qubit model emerges at this point: there are crosses within the spectra of each parity subspace but for exceptional cases like that treated in \cite{Peng2012p365302}.
These crossings in the spectra point to classify the system as not integrable for the majority of the parameter sets \cite{Braak2011p100401}.  
Then, we present the proper functions via linear algebra methods and Bargmann representation \cite{Bargmann1947p568}.
In the second to last section, we show how the time evolution in the weak-coupling regime can be calculated in closed form and how to calculate the evolution of quantities of interest in the parity decomposition. 
Finally, we close with a brief conclusion.

%
\section{Eigenvalues of the model}

It is trivial to calculate the eigenvalues of the general two-qubit quantum Rabi model (\ref{eq:Hamiltonian}) in the weak-coupling regime, $g_{1}, g_{2} \ll \omega_{f}$, where the RWA approximation is valid; i.e. the term $\sum_{j=1,2} \left( \hat{a} + \hat{a}^{\dagger} \right) g_{j} \hat{\sigma}_{x}^{(j)} $  becomes  $\sum_{j=1,2} g_{j} \left( \hat{a} \hat{\sigma}_{+}^{(j)} +\hat{a}^{\dagger} \hat{\sigma}_{-}^{(j)} \right)$. 
For the Dicke model, $\omega_{1} = \omega_{2} = \omega$ and $g_{1}= g_{2} = g$, it is safe to say that the exact eigenvalues in the RWA describes well the system when the relation between the coupling strengths and the field frequency is below $10\%$, $g_{j} / \omega_{f} < 0.1$ \cite{Wang2012p014031}. 
The same can be said for the majority of cases but we find that the eigenvalues provided by the rotating wave approximation for one coupling similar or larger than $10\%$ of the field frequency can be within $1\%$ of the numerical values for the full Hamiltonian by restricting the other coupling well below $10\%$ of the field frequency.
Something similar occurs via symmetric detuning of the qubits with respect to the field frequency, the Hamiltonian under the RWA provides eigenvalues in good agreement with the full Hamiltonian for couplings larger than $10\%$ of the field frequency; e.g. for symmetric detunings $\omega_{1}= \omega_{f} - \Delta $ and $\omega_{2}= \omega_{f} + \Delta $ with $\Delta \in [0.1,0.5]$, the ground energies provided by the Hamiltonian under the RWA are within $2.4\%$ of those provided by the full Hamiltonian for coupling values of up to $20\%$ of the field frequency, $g_{1}=g_{2}= 0.2 ~\omega_{f}$, and the mean relative error for the first twenty eigenvalues provided by the RWA Hamiltonian compared to the full Hamiltonian is up to $7.5\%$.  
 
We can also approximate the eigenvalues in the DSC regime by assuming that the coupling parameters are larger than the transition frequencies, $g_{1}, g_{2} \gg \omega_{1}, \omega_{2}$.
In such a case, we can write the two-qubit quantum Rabi model (\ref{eq:Hamiltonian}) as a leading Hamiltonian with a perturbation, $\hat{H} = \hat{H}_{0} + \hat{P}$, where we can use the general rotation $\hat{R}_{y} = e^{-i \hat{\sigma}_{y}^{(1)} \theta} \otimes e^{-i \hat{\sigma}_{y}^{(2)} \theta}$ with $\theta = \pi/4$ to implement the changes $\hat{\sigma}_{x}^{(j)} \rightarrow \hat{\sigma}_{z}^{(j)}$ and $\hat{\sigma}_{z}^{(j)} \rightarrow -\hat{\sigma}_{x}^{(j)}$ leading to the form,
\begin{eqnarray}
\tilde{H}_{0} &=& \omega_{f} \hat{n} +  \left( \hat{a} + \hat{a}^{\dagger} \right) \left( g_{1} \hat{\sigma}_{z}^{(1)} + g_{2}\hat{\sigma}_{z}^{(2)}\right), \\
\tilde{P} &=& -\frac{1}{2} \left( \omega_{1} \hat{\sigma}_{x}^{(1)} + \omega_{2} \hat{\sigma}_{x}^{(2)}  \right),
\end{eqnarray}
where the tilde has been used to represent the rotated operator, $\tilde{O} = \hat{R}^{\dagger}_{y}(\pi/4) \hat{O} \hat{R}_{y}(\pi/4)$.
The unperturbed Hamiltonian $\tilde{H}_{0}$ is diagonal,
\begin{eqnarray}
\tilde{H}_{0} = \hat{T}_{D} \left[  \omega_{f} \hat{n} - \frac{1}{\omega_{f}} \left( \begin{array}{cccc} 
g_{+}^{2}&0&0&0 \\
0&g_{-}^{2}&0&0 \\
0&0&g_{-}^{2}&0 \\
0&0&0&g_{+}^{2} \\
\end{array}
\right)
\right] \hat{T}_{D}^{\dagger},
\end{eqnarray}
in a driven oscillator basis provided by 
\begin{eqnarray}
\hat{T}_{D} = \left( \begin{array}{cccc} 
\hat{D}(g_{+}/\omega_{f})&0&0&0 \\
0&\hat{D}(g_{-}/\omega_{f})&0&0 \\
0&0&\hat{D}(-g_{-}/\omega_{f})&0 \\
0&0&0&\hat{D}(-g_{+}/\omega_{f}) \\
\end{array}
\right).
\end{eqnarray}
The displacement operator is defined as $\hat{D}(\alpha) = e^{\alpha \hat{a}^{\dagger} - \alpha^{\ast} \hat{a}}$ and the auxiliary coupling parameters as $g_{\pm} = g_{1} \pm g_{2}$.
Thus, the approximated eigenvalues up to second order correction in the DSC regime are twofold degenerate:
\begin{eqnarray}
\epsilon_{1,m} &=& \epsilon_{4,m} \approx \omega_{f} m - \frac{g_{+}^{2}}{\omega_{f}} - \epsilon_{2,+}, \label{eq:SpBranch1}\\
\epsilon_{2,m} &=& \epsilon_{3,m} \approx \omega_{f} m - \frac{g_{-}^{2}}{\omega_{f}} - \epsilon_{2,-}, \label{eq:SpBranch2}
\end{eqnarray}
where the second order correction is given by
\begin{eqnarray}
\epsilon_{2,\pm} &=& \sum_{m\ne n} \frac{\omega_{1}^{2} \vert \langle m \vert \hat{D}(2g_{1}/\omega_{f}) \vert n \rangle \vert^{2}}{\omega_{f} (m-n) \pm 4 g_{1} g_{2} /\omega_{f} } + \sum_{m\ne n} \frac{\omega_{2}^{2} \vert \langle m \vert \hat{D}(2g_{2}/\omega_{f}) \vert n \rangle \vert^{2}}{\omega_{f} (m-n) \pm 4 g_{1} g_{2} /\omega_{f} },
\end{eqnarray}
alongside the the identity  $\langle m \vert \hat{D}(2x) \vert n \rangle = \sqrt{\frac{m!}{n!}} (2x)^{(m-n)} e^{-2 \vert x\vert^{2}} \mathcal{L}_{n}^{(m-n)}(\vert 2 x \vert^{2})$ where $\mathcal{L}_{a}^{(b)}(z)$ is a generalized Laguerre polynomial. 
The first order correction due to the perturbation $\tilde{P}$ is null. 

\begin{figure}
\center\includegraphics[scale=1]{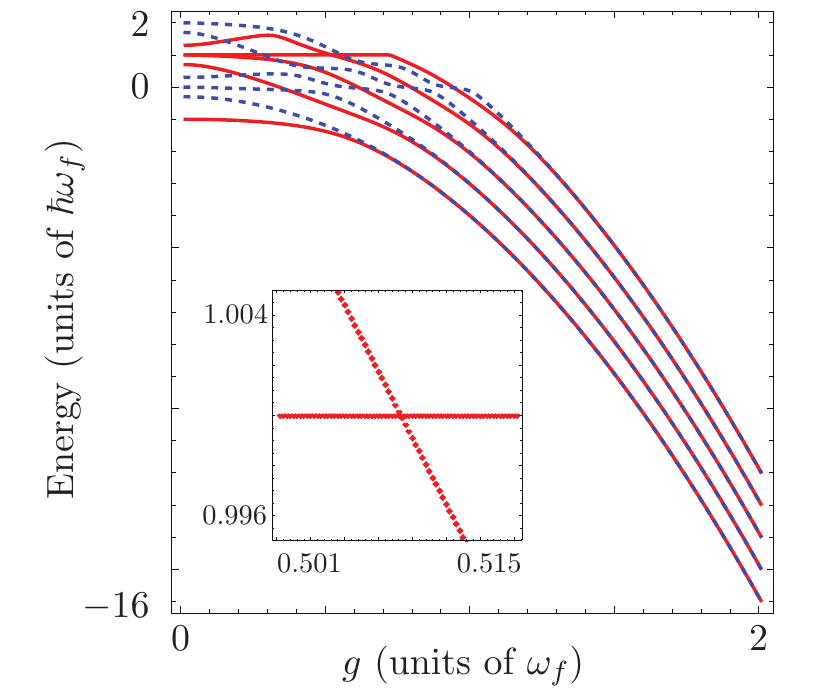}
\caption{(Color online) A piece of the numerical spectra for the quantum Rabi Hamiltonian considering subspaces of up to one thousand excitations for even (solid red) and odd (dashed blue) parity subspaces for the parameter sets: $\{\omega_{1}, \omega_{2}, g_{1}, g_{2} \} = \{ 1.3 , 0.7, g, g \}  \omega_{f}$. The inset shows a crossing in the even parity spectra of the system at hand. The driven oscillator behavior of the spectra, predicted by perturbation theory can already be seen for couplings twice as large as the transition frequencies for the lower eigenvalues.}
\label{fig:Fig1}
\end{figure}

Note that for  extremely large values of the coupling parameters $g_{1}$ and $g_{2}$, i.e. well into the DSC regime, the second-order corrections tend to zero and each of the spectral branches of the two-qubit quantum Rabi model, (\ref{eq:SpBranch1}) and (\ref{eq:SpBranch2}), tends to the spectra of a forced oscillator. 
Figure \ref{fig:Fig1} shows the numerical spectra for the quantum Rabi Hamiltonian expanded in the parity subspaces spanned by the even, (\ref{eq:EvenChain}), and odd, (\ref{eq:OddChain}), parity bases defined in the following section.
We consider up to 1000 excitations in the system; i.e. truncated $H_{\pm}$ matrices given by (\ref{eq:HpmMatrix}) of size 1000 with coupling steps of one-hundredth of the field frequency, $\Delta g = 0.01~\omega_{f}$.
We show the symmetric off-resonance case, $\omega_{1}= 1.3 ~\omega_{f}$ and $\omega_{2}= 0.7 ~\omega_{f}$, with identical couplings, $g_{1} = g_{2} = g \in [0,2 ]\omega_{f} $.
The spectra corresponding to even parity eigenvectors are presented as solid red lines and that corresponding to odd parity are shown as dashed blue lines. 
The inset in Fig. \ref{fig:Fig1} shows an energy crossing in the even spectra confirmed by comparison of the proper states before and after the crossing. 

We studied a variety of cases not shown here; e.g. random qubit transition energies $\omega_{1}, \omega_{2} \in [0,2]\omega_{f}$ with identical couplings $g_{1}= g_{2}=g$ or one fixed random coupling $g_{1} \in [0,2]\omega_{f}$. 
In all the cases studied for nonidentical qubits, energy crossings in the spectra of each parity subspaces were the norm and the spectra of the parity subspaces always cross each other.
The fact that energy crossings appear in the parity subspaces remarks the deviation from the expected behavior for the single-qubit quantum Rabi model \cite{Braak2011p100401} where there lack of energy crossings in the spectra of parity subspaces allows each eigenstate  to be uniquely labeled by the excitation number and parity; the ability to label each eigenstate uniquely is regarded as a criterion of quantum integrability in \cite{Braak2011p100401}.
Also, in the cases of identical couplings, $g_{1}= g_{2}=g$, the lower eigenvalue branches already tend to the approximated spectra of two driven oscillator chains for not-so-large values of the coupling parameters; in the case shown in Fig. \ref{fig:Fig1} the first dozen eigenvalue branches already show this behavior for $g_{1} = g_{2} \ge 2  \omega_{f}$.

%
\section{Eigenstates of the model}

Here we calculate the proper states of the system. 
For the sake of completeness we use both a linear algebra and the Bargmann representation method that lead to four- and five-term recurrence relations between the coefficients of the eigenstates, respectively.
The eigenstates provided by both methods are valid for any given coupling parameter value.

\subsection{Linear algebra approach} \label{sec:LAP}
 
The fact that the model conserves parity allows us to extend the idea of a parity basis used in the single-qubit Rabi model \cite{Tur2000p574, Tur2001p899, Casanova2010p263603},
\begin{eqnarray}
\left\{ \vert j \rangle_{+} \right\} &=& \left\{ \vert 0,g, g \rangle, \vert 0,e, e \rangle, \vert 1,e, g \rangle, \vert 1,g, e \rangle,\vert 2,g, g \rangle, \vert 2,e, e \rangle,\ldots \right\}, \label{eq:EvenChain}\\
\left\{ \vert j \rangle_{-} \right\} &=& \left\{ \vert 0,e, g \rangle, \vert 0,g, e \rangle, \vert 1,g, g \rangle, \vert 1,e, e \rangle,\vert 2,e,g \rangle, \vert 2,g, e \rangle,\ldots \right\}, \label{eq:OddChain}
\end{eqnarray}
with $\hat{\Pi} \vert j \rangle_{\pm} = \pm \vert j \rangle_{\pm}$.
The model Hamiltonian (\ref{eq:Hamiltonian}) in these bases, $(H_{\pm} )_{j,k} = _{\pm}\langle j \vert \hat{H} \vert k \rangle_{\pm}$, has block tridiagonal form:
\begin{eqnarray} \label{eq:HpmMatrix}
H_{\pm} = \left( \begin{array}{ccccc}
D_{0}^{\pm} & O_{1} & 0 & 0 &\ldots \\
O_{1} & D_{1}^{\pm} & O_{2} &0 & \ldots \\
0& O_{2} & D_{2}^{\pm} & O_{3} & \ldots \\
\vdots &\vdots &\vdots &\vdots &\vdots 
\end{array} \right),
\end{eqnarray}
where the blocks are given by
\begin{eqnarray}
D_{j}^{\pm} = \left( \begin{array}{cc} d^{+}_{\pm}
 &0 \\
0&  d^{-}_{\pm}
\end{array} \right), \quad O_{j} = \sqrt{j} \left( \begin{array}{cc}
g_{1} & g_{2} \\ g_{2} & g_{1}
\end{array} \right).
\end{eqnarray}
with $ d^{+}_{\pm} = j \omega_{f}  \mp \frac{1}{2}\left[ (-1)^{j} \omega_{1} \pm \omega_{2}\right]$ and $ d^{-}_{\pm} = j \omega_{f} \pm \frac{1}{2}\left[  (-1)^{j}  \omega_{1} \pm  \omega_{2}\right]$.
Notice that the matrices $O_{j}$ are invertible as long as $\vert g_{1} \vert^{2} \ne \vert g_{2} \vert^{2} $.
Thus, the eigenstates are written as
\begin{eqnarray}
\vert \xi_{\pm} \rangle &\propto& \sum_{j=0}^{\infty} v_{j,0}^{(\pm)} \vert \pm,j,g \rangle + v_{j,1}^{(\pm)} \vert \mp,j,e \rangle,
\end{eqnarray} 
where we have used the symbol $\xi_{\pm}$ to emphasize that the proper functions delivered by this method are valid for any given coupling parameter values.
The coefficients can be calculated up to a normalization factor by the four-term recurrence relations given by
\begin{eqnarray}
\vec{v}_{1}^{(\pm)} &=& - O_{1}^{-1} \left( D_{0}^{\pm}  - \mathbbm{1} \xi_{\pm} \right) \vec{v}_{0}^{(\pm)}, \label{eq:RRLA1}\\
\vec{v}_{j}^{(\pm)} &=& - O_{j}^{-1} \left( D_{j-1}^{\pm}  - \mathbbm{1} \xi_{\pm} \right) \vec{v}_{j-1}^{(\pm)} - O_{j}^{-1} O_{j-1} \vec{v}_{j-2}^{(\pm)}, \quad j=2,3,\ldots \label{eq:RRLA2}
\end{eqnarray}
by choosing a suitable $\vec{v}_{0}^{(\pm)}$, with $\vec{v}_{j}^{(\pm)} = (v_{j,0}^{(\pm)}, v_{j,1}^{(\pm)})$ and using 
\begin{eqnarray}
O_{j}^{-1} \left( D_{j-1}^{\pm}  - \mathbbm{1} \xi_{\pm} \right)&=& \frac{1}{\sqrt{j} (g_{1}^{2} - g_{2}^{2})}  \left( \begin{array}{cc}
g_{1} (d^{+}_{\pm} - \xi_{\pm}) & - g_{2} (d^{-}_{\pm} -  \xi_{\pm}) \\
-g_{2} (d^{+}_{\pm}- \xi_{\pm}) & g_{1} (d^{+}_{\pm}- \xi_{\pm})
\end{array} \right) , \nonumber \\~\\
O_{j}^{-1} O_{j-1} &=& \sqrt{\frac{j-1}{j}} ~\mathbbm{1},
\end{eqnarray}
where the symbol $\mathbbm{1}$ is the unitary matrix of dimension two. 

We want to emphasize that keeping track of the relative error is of great importance when calculating the eigenstates via the recurrence relations given in (\ref{eq:RRLA1}) and (\ref{eq:RRLA2}).
A safer approach is to utilize standard linear algebra diagonalization packages for large sparse matrices.

\subsection{Bargmann representation method}

We can follow the procedure presented in \cite{Peng2012p365302} with our Hamiltonian model (\ref{eq:Hamiltonian}).
That is, we use the transformation defined above and, also, move into the basis defined by $\hat{\sigma}_{z}^{(1)}$. 
Then, we use the  Bargmann representation method, i.e., making the substitution $\hat{a} \rightarrow \partial_{z}$ and $\hat{a}^{\dagger} \rightarrow z $ where the shorthand notation $\partial_{x}$ has been used to denote partial derivations with respect to $x$, to arrive to the Hamiltonian 
\begin{eqnarray}
\hat{H}_{R} = \left( \begin{array}{cc}
\omega_{f} z \partial_{z} - \frac{\omega_{2}}{2} \sigma_{x}^{(2)} + (g_{2} \sigma_{z}^{(2)} + g_{1} )(z + \partial_{z}) & - \frac{\omega_{1}}{2} \\
 - \frac{\omega_{1}}{2} & \omega_{f} z \partial_{z} - \frac{\omega_{2}}{2} \sigma_{x}^{(2)} + (g_{2} \sigma_{z}^{(2)} - g_{1} )(z + \partial_{z}) 
\end{array} \right). \nonumber \\
\end{eqnarray}
At this point, we can use the so-called Fulton-Gouterman transformation \cite{Fulton1961p1059, Braak2011p100401, Peng2012p365302} to reduce the eigenvalue problem to the form of four coupled differential equations,
\begin{eqnarray}
\left[ \omega_{f} z \partial_{z} + g_{+}(z + \partial_{z}) - \chi_{\pm} \right] \phi_{1} ^{\pm} - \frac{\omega_{2}}{2} \phi_{2}^{\pm} \pm \frac{\omega_{1}}{2} \bar{\phi}_{2}^{\pm} = 0, \\
\left[ \omega_{f} z \partial_{z}  + g_{-}(z + \partial_{z})  - \chi_{\pm} \right] \phi_{2}^{\pm}  - \frac{\omega_{2}}{2} \phi_{1}^{\pm} \pm \frac{ \omega_{1}}{2} \bar{\phi}_{1}^{\pm} = 0, \\
\left[ \omega_{f} z \partial_{z}  - g_{+}(z + \partial_{z}) - \chi_{\pm} \right] \bar{\phi}_{1}^{\pm}  - \frac{ \omega_{2}}{2} \bar{\phi}_{2}^{\pm} \pm \frac{ \omega_{1}}{2} \phi_{2}^{\pm} = 0, \\
\left[ \omega_{f} z \partial_{z} - g_{-}(z + \partial_{z}) - \chi_{\pm} \right] \bar{\phi}_{2}^{\pm}  - \frac{ \omega_{2}}{2} \bar{\phi}_{1}^{\pm} \pm \frac{\omega_{1}}{2} \phi_{1}^{\pm} = 0, 
\end{eqnarray}
where we have used the shorthand notation $\phi_{j}^{\pm} \equiv \phi_{j}^{\pm}(z)$ and $\bar{\phi}_{j}^{\pm} = \phi_{j}^{\pm}(-z)$.
Here, we introduce a deviation from previous work and take advantage of parity.
We use a pair of Bargmann functions with well defined parity, $\Phi_{j,\pm}^{\pm} = \phi_{j}^{\pm} \pm \bar{\phi}_{j}^{\pm}$ such that $\hat{T}_{FG} \Phi_{j,\pm}^{\pm} = \pm \Phi_{j,\pm}^{\pm}$, that yield a coupled differential set,
\begin{eqnarray}
\left[ \omega_{f} z \partial_{z} - \chi_{\pm} \right] \Phi_{1,+}^{\pm} + g_{+}(z + \partial_{z}) \Phi_{1,-}^{\pm}  - \frac{\omega_{2} \mp \omega_{1}}{2} \Phi_{2,+}^{\pm} = 0, \label{eq:Phi2plus} \\
\left[ \omega_{f} z \partial_{z} -\chi_{\pm} \right] \Phi_{2,+}^{\pm} + g_{-}(z + \partial_{z}) \Phi_{2,-}^{\pm}  - \frac{\omega_{2} \mp \omega_{1}}{2} \Phi_{1,+}^{\pm} = 0, \\ 
\left[ \omega_{f} z \partial_{z} - \chi_{\pm} \right] \Phi_{1,-}^{\pm} + g_{+}(z + \partial_{z}) \Phi_{1,+}^{\pm}  - \frac{\omega_{2} \pm \omega_{1}}{2} \Phi_{2,-}^{\pm} = 0,  \label{eq:Phi2minus}\\
\left[ \omega_{f} z \partial_{z} - \chi_{\pm} \right] \Phi_{2,-}^{\pm} + g_{-}(z + \partial_{z}) \Phi_{2,+}^{\pm}  - \frac{\omega_{2} \pm \omega_{1}}{2} \Phi_{1,-}^{\pm} = 0,
\end{eqnarray}
that can be solved by Frobenius method with the power series solutions, 
\begin{eqnarray}
\Phi_{1,+}^{\pm} = \sum_{k=0}^{\infty} c_{2k}^{\pm} z^{2k}, \qquad
\Phi_{1,-}^{\pm} = \sum_{k=0}^{\infty} c_{2k+1}^{\pm} z^{2k+1},
\end{eqnarray}
for one of the pairs of even and odd parity Bargmann functions; the coefficients for the other pair, $\Phi_{2,\pm}^{\pm}$, are obtained as functions of the coefficients $c_{k}^{\pm}$ from (\ref{eq:Phi2plus}) and (\ref{eq:Phi2minus}) .
Such an approach yields a five-term recurrence relation for the coefficients $c_{k}^{\pm}$,
\begin{eqnarray}
\alpha_{0}^{\pm}(2) c_{2}^{\pm} + \alpha_{1}^{\pm}(2) c_{1}^{\pm} + \alpha_{2}^{\pm}(2) c_{0}^{\pm} = 0,   \\
\alpha_{0}^{\pm}(3) c_{3}^{\pm} + \alpha_{1}^{\pm}(3) c_{2}^{\pm} + \alpha_{2}^{\pm}(3) c_{1}^{\pm} + \alpha_{3}^{\pm}(3) c_{0}^{\pm} = 0,   \\
\alpha_{0}^{\pm}(j) c_{j}^{\pm} + \alpha_{1}^{\pm}(j) c_{j-1}^{\pm} + \alpha_{2}^{\pm}(j) c_{j-2}^{\pm} + \alpha_{3}^{\pm}(j) c_{j-3}^{\pm} + \alpha_{4}^{\pm}(j) c_{j-4}^{\pm} = 0, \quad j=4,5,\ldots \nonumber \\
\end{eqnarray}
with
\begin{eqnarray}
\alpha_{0}^{\pm}(j) &=& \frac{j!}{(j-2)!} ~g_{+}g_{-} \left[ \omega_{1} \mp (-1)^{j} \omega_{2} \right], \\
\alpha_{1}^{\pm}(j) &=& \left( j-1 \right) \left\{ g_{-} \left[ \omega_{1} \mp (-1)^{j} \omega_{2} \right] \left[ (j-1)\omega_{f} -\chi_{\pm} \right] \right. + \nonumber \\
&& \left. +  g_{+} \left[ \omega_{1} \pm (-1)^{j} \omega_{2} \right] \left[ \chi_{\pm} -(j-2)\omega_{f} \right] \right\}, \\
\alpha_{2}^{\pm}(j) &=&  (2j-3) g_{+} g_{-} \left[ \omega_{1} \mp (-1)^{j} \omega_{2} \right] + \left[ \omega_{1} \pm (-1)^{j} \omega_{2} \right] \times \nonumber \\
&& \times \left\{\frac{1}{4}\left[\omega_{1} \mp (-1)^{j} \omega_{2} \right]^{2} - \left[ \chi_{\pm} - \left(j - 2 \right)\omega_{f} \right]^{2} \right\},\\
\alpha_{3}^{\pm}(j) &=& g_{+} \left[ \omega_{1} \pm (-1)^{j} \omega_{2} \right] \left[ \chi_{\pm} - (j-2) \omega_{f}\right] -\nonumber \\
&& - g_{-} \left[ \omega_{1} \mp (-1)^{j} \omega_{2} \right] \left[ \chi_{\pm} -(j-3)\omega_{f} \right], \\
\alpha_{4}^{\pm}(j) &=& g_{+} g_{-} \left[ \omega_{1} \mp (-1)^{j} \omega_{2} \right].
\end{eqnarray}

For identical qubits, $\omega_{1} = \omega_{2} = \omega_{0}$, the differential set for the $H_{+}$ block,
\begin{eqnarray}
(\omega_{f} z \partial_{z} - \chi_{+} ) \Phi_{1,+}^{+} + g_{+} (z + \partial_{z}) \Phi_{1,-}^{+} &=& 0, \\
(\omega_{f} z \partial_{z} -\chi_{+} ) \Phi_{2,-}^{+} - \omega_{0} \Phi_{1,-}^{+} &=& 0, \\
(\omega_{f} z \partial_{z} -\chi_{+} ) \Phi_{1,-}^{+} + g_{+}(z + \partial_{z})\Phi_{1,+}^{+} - \omega_{0} \Phi_{2,-}^{+} &=& 0, \label{eq:Odd2} \\
(\omega_{f} z \partial_{z} - \chi_{+}) \Phi_{2,+}^{+}  &=& 0, \label{eq:Even2}
\end{eqnarray}
immediately shows us that the Bargman function $\phi_{2}^{+}(z)$ has well defined odd parity as Eq.(\ref{eq:Even2}) accepts only the trivial solution $\Phi_{2,+}^{+}(z) = 0$. 
In this case the five-term recurrence relations reduce to three-term recurrence relations for the coefficients, where we have rearranged the coefficients for the sake of simplicity,
\begin{eqnarray}
c_{j}^{\pm} &=& \frac{1}{j} \left[ \alpha_{j}^{\pm} c_{j-1}^{\pm} - \left(1 - \delta_{1,j}\right) c_{j-2}^{\pm}  \right], \quad j = 1, 2, \ldots \\
\alpha_{j}^{\pm} &=&   \frac{[\chi_{\pm}-(j-1) \omega_{f}]^2 - \left[\frac{1 \pm (-1)^{j}}{2} \right]^{2} \omega_{0}^2 }{g_{+}[\chi_{\pm} - (j-1) \omega_{f}]}.  
\end{eqnarray}
Negative superindex corresponds to the differential set for the $H_{-}$ block,
\begin{eqnarray}
(\omega_{f} z \partial_{z} - \chi_{-}) \Phi_{1,-}^{-} + g_{+} (z + \partial_{z}) \Phi_{1,+}^{-}  &=& 0, \\
(\omega_{f} z \partial_{z} - \chi_{-}) \Phi_{2,+}^{-} -  \omega_{0} \Phi_{1,+}^{-} &=& 0, \\
(\omega_{f} z \partial_{z} -\chi_{-} ) \Phi_{1,+}^{-} + g_{+}(z + \partial_{z})\Phi_{1,-}^{-} - \omega_{0} \Phi_{2,+}^{-}  &=& 0, \\
(\omega_{f} z \partial_{z} - \chi_{-}) \Phi_{2,-}^{-}  &=& 0.
\end{eqnarray}
with $\Phi_{1,+}^{\pm} = \sum_{j=0}^{\infty} c_{2j}^{\pm} z^{2j}$ and $\Phi_{1,-}^{\pm} = \sum_{j=0}^{\infty} c_{2j+1}^{\pm} z^{2j+1}$.
The coefficients in the three-term recurrence relations satisfy $
\lim_{j \rightarrow \infty } \frac{\alpha_{j}^{\pm}}{\alpha_{j+1}^{\pm}} =  1$.

%
\section{Time Evolution}

We want to include a way of calculating the exact time evolution in the RWA and show that numerical solutions for the full quantum problem are simplified by using the parity bases introduced in (\ref{eq:EvenChain}) and (\ref{eq:OddChain}).
In order to round up our example we show some typical measurements on radiation-matter interaction systems; e.g., mean photon number, $\hat{n}$, population inversion, $\hat{S}_{z} = (\hat{\sigma}_{z}^{(1)} + \hat{\sigma}_{z}^{(2)})/2$, von Neumann entropy of the bipartite system, $\hat{S} = - \hat{\rho}_{q} \ln \hat{\rho}_{q}$, and concurrence as defined by Wootters \cite{Wootters1998p2245, Wootters2001p27}.

\subsection{Weak-coupling}

Let us start from the Tavis-Cummings Hamiltonian for two-qubits in the frame defined by the transformation $\hat{U}_{\hat{N}}(t)= e^{- i \omega_{f} \hat{N} t}$,
\begin{eqnarray}
\hat{H}_{RWA} = \sum_{j=1,2} \Delta_{j} \hat{\sigma}_{z}^{(j)} + \sum_{j=1,2} g_{j} \left( \hat{a} \hat{\sigma}_{+}^{(j)} + \hat{a}^{\dagger} \hat{\sigma}_{-}^{(j)} \right),
\end{eqnarray}
where the detunnings are given by $\Delta_{j} = \left( \omega_{j} - \omega_{f}\right)/2$.
One of us has pointed out somewhere else \cite{RodriguezLara2013p095301} that the right unitary transformation,
\begin{eqnarray}
\hat{T}_{j} = \left( \begin{array}{cc} \hat{V}&0\\0&1 \end{array} \right)_{j},
\end{eqnarray}
diagonalizes the single-qubit Jaynes-Cummings model in the field basis.
So, one can use that transformation for each qubit and obtain,
\begin{eqnarray}
\left(\hat{T}_{1} \otimes \hat{T}_{2}\right) \hat{H}_{sc}  \left( \hat{T}_{2}^{\dagger} \otimes \hat{T}_{1}^{\dagger} \right) = \hat{H}_{RWA},
\end{eqnarray}
with 
\begin{eqnarray}
\hat{H}_{sc} = \left( 
\begin{array}{cccc} 
\Delta_{1}+\Delta_{2}& g_{2}\sqrt{n-1}&g_{1}\sqrt{n-1}&0 \\ 
g_{2}\sqrt{n-1} & \Delta_{1}-\Delta_{2}& 0 &g_{1}\sqrt{n} \\ 
g_{1}\sqrt{n-1}&0&-\Delta_{1}+\Delta_{2}&g_{2}\sqrt{n} \\ 
0&g_{1}\sqrt{n} & g_{2}\sqrt{n}&-\Delta_{1}-\Delta_{2} \\ 
\end{array} \right).
\end{eqnarray}
It can be shown that the corresponding evolution operator of the system in the RWA is
\begin{eqnarray}
\hat{U}_{RWA}(t) = \left(\hat{ T}_{1} \otimes \hat{T}_{2}\right) e^{-i \hat{H}_{sc} t} \left( \hat{T}_{2}^{\dagger} \otimes \hat{T}_{1}^{\dagger} \right). \label{eq:URWA}
\end{eqnarray}
This requires just to keep track of the action of $\left(\hat{T}_{2}^{\dagger} \otimes \hat{T}_{1}^{\dagger} \right)$ over the given initial state in order to calculate its evolution.
It is simple to calculate the exponential because the Hamiltonian $\hat{H}_{sc}$ has a depressed quartic as characteristic polynomial,
\begin{eqnarray}
\vert H_{sc} - \lambda I \vert &=& c_{0} + c_{1} \lambda + c_{2} \lambda^2 + \lambda^4, \\
c_{0}&=&\left[ n (g_{1}^2 - g_{2}^2) + \Delta_{1}^{2} - \Delta_{2}^{2}\right]\left[ (n-1) (g_{1}^2 - g_{2}^2) + \Delta_{1}^{2} - \Delta_{2}^{2} \right], \\
c_{1}&=& 2 \left( g_{2}^{2} \Delta_{1} + g_{1}^{2} \Delta_{2} \right),\\
c_{2} &=& (1-2n)(g_{1}^2 + g_{2}^2) - 2 \left( \Delta_{1}^{2} + \Delta_{2}^{2} \right),
\end{eqnarray}
with roots, 
\begin{eqnarray}
\lambda_{1,2}&=& \frac{ p \pm \sqrt{ - p^2 - 2 c_{2} - 2 c_{1}/p}}{2 }, \\
\lambda_{3,4}&=& \frac{- p \pm \sqrt{ - p^2 - 2 c_{2} + 2 c_{1}/p}}{2}, 
\end{eqnarray}
and parameters,
\begin{eqnarray}
p &=& \sqrt{ \frac{12 c_{0} + (c_{2} -c)^2}{3 c}},\\
c &=& \left[\frac{1}{2} (q + 27 c_{1}^2 - 72 c_{0} c_{2} + 2 c_{2}^3)\right]^{1/3}, \\
q&=& \sqrt{ (27 c_{1}^2 - 72 c_{0} c_{2} + 2 c_{2}^3)^2 - 4 (12 c_{0} + c_{2}^{2})^3}.
\end{eqnarray}
Once the eigenvalues are obtained, it is trivial but cumbersome to calculate the eigenstates of $\hat{H}_{sc}$ that allow us to write the evolution operator $\hat{U}(t)$.

\subsection{Parity based numeric approach}

\begin{figure}
\center\includegraphics[scale=1]{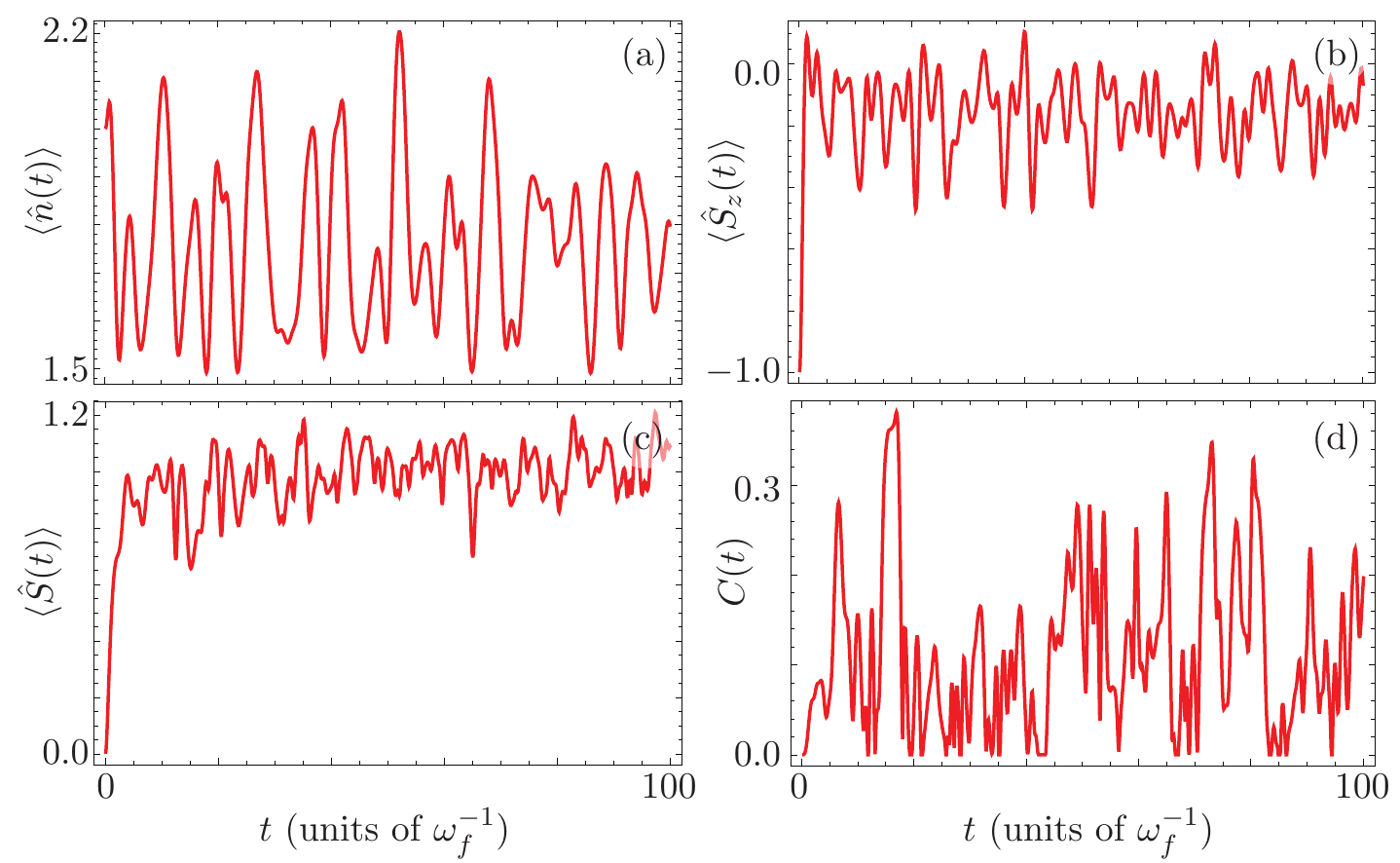}
\caption{(Color online) The time evolution of the (a) mean photon number, (b) population inversion, (c) von Neuman entropy and (d) bipartite concurrence for a quantum Rabi Hamiltonian  with the parameter set $\{\omega_{1}, \omega_{2}, g_{1}, g_{2} \} = \{ 1.1 , 0.3, 0.3, 0.4 \} \omega_{f}$ and initial state $\vert \phi(0) \rangle = \vert \alpha, g, g \rangle$ with the coherent field parameter $\alpha = \sqrt{2}$  considering a subspace of up to one thousand excitations.}
\label{fig:Fig2}
\end{figure}

\begin{figure}
\center\includegraphics[scale=1]{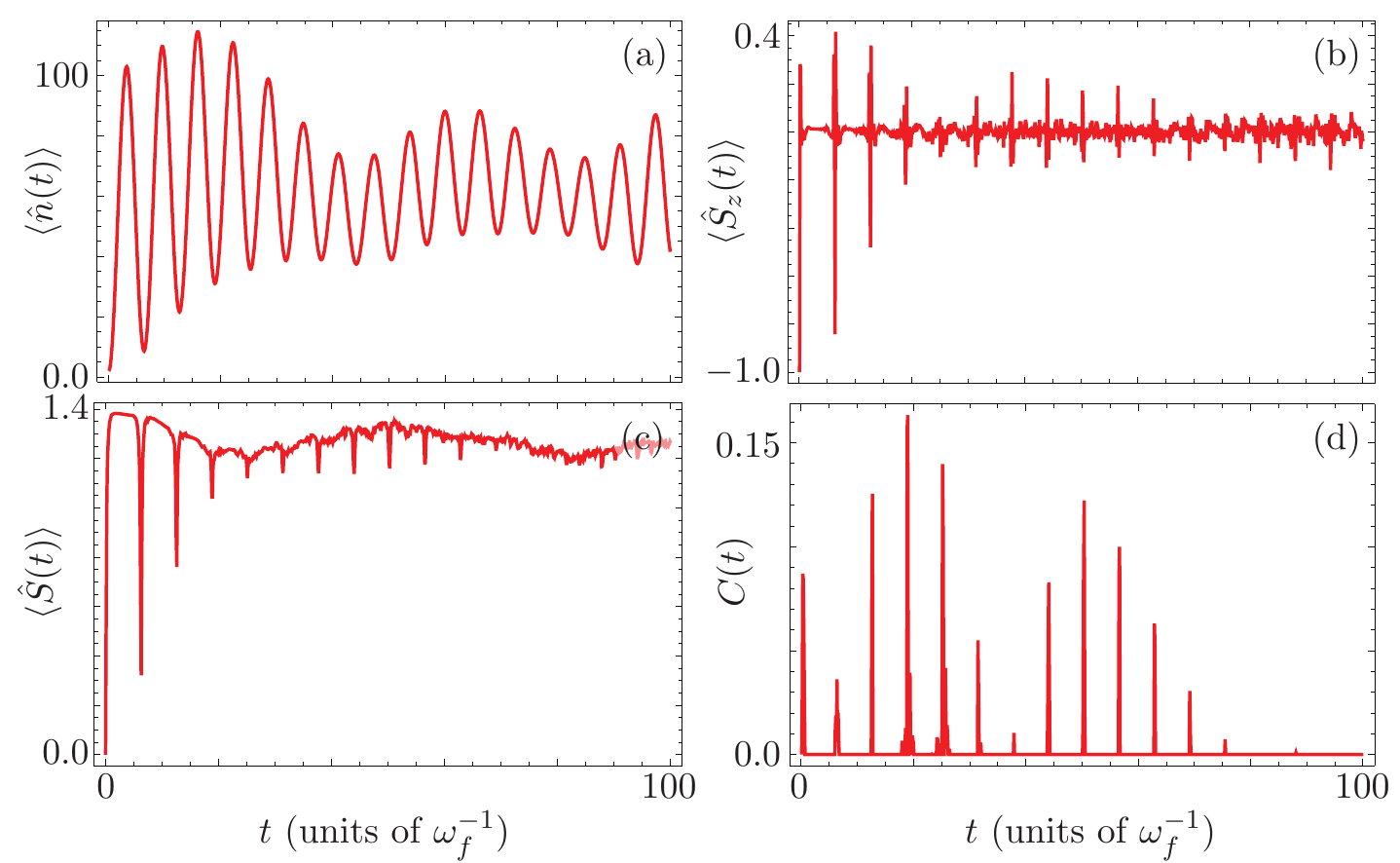}
\caption{(Color online) The time evolution of the (a) mean photon number, (b) population inversion, (c) von Neuman entropy and (d) bipartite concurrence for a quantum Rabi Hamiltonian  with the parameter set $\{\omega_{1}, \omega_{2}, g_{1}, g_{2} \} = \{ 1.1 , 0.3, 3, 4 \} \omega_{f}$ and initial state $\vert \phi(0) \rangle = \vert \alpha, g, g \rangle$ with the coherent field parameter $\alpha = \sqrt{2}$  considering a subspace of up to one thousand excitations.}
\label{fig:Fig3}
\end{figure}

In the most general case of any given coupling parameter set, one can approximate results by adequately truncating the ${H}_{\pm}$ matrices in (\ref{eq:HpmMatrix}) obtained by using the parity bases (\ref{eq:EvenChain}) and (\ref{eq:OddChain}).
This numerical approach allows us to calculate the time evolution of a general initial state decomposed in the even and odd parity chain bases, respectively, 
\begin{eqnarray}
\vert \psi(0) \rangle_{+} &=& \sum_{n=0}^{\infty} c_{4n}^{(+)} \vert 2n, g, g\rangle + c_{4n+1}^{(+)} \vert 2n, e, e\rangle + \nonumber \\
&&+c_{4n+2}^{(+)} \vert 2n+1, e, g\rangle + c_{4n+3}^{(+)} \vert 2n+1, g, e\rangle , \\
\vert \psi(0) \rangle_{-} &=& \sum_{n=0}^{\infty} c_{4n}^{(-)} \vert 2n, e, g\rangle + c_{4n+1}^{(-)} \vert 2n, g, e\rangle + \nonumber \\
&&+c_{4n+2}^{(-)} \vert 2n+1, g, g\rangle + c_{4n+3}^{(-)} \vert 2n+1, e, e\rangle
\end{eqnarray} 
with the prescription, $ \vec{c}^{(\pm)}(t) = e^{- i \hat{H}_{\pm}  t} \vec{c}^{(\pm)}(0)$, where the notation $\vec{c}^{(\pm)}$ corresponds to a vector of state amplitudes.
Any quantity of interest can be calculated from the state above or the qubit ensemble reduced density matrix, 
\begin{eqnarray}
\hat{\rho}_{q} &=&\sum_{n=0}^{\infty} \left( \begin{array}{cccc}
\vert c_{4n+1}^{(+)}\vert^2  & c_{4n+1}^{(+)}c_{4n}^{(-)\ast}   & c_{4n+1}^{(+)}c_{4n+1}^{(-)\ast}  &  c_{4n+1}^{(+)}c_{4n}^{(+)\ast} \\
c_{4n+2}^{(+)}c_{4n+3}^{(-)\ast} & \vert c_{4n+2}^{(+)}\vert^2 &c_{4n+2}^{(+)}c_{4n+3}^{(+)\ast}&c_{4n+2}^{(+)}c_{4n+2}^{(-)\ast} \\
c_{4n+3}^{(+)}c_{4n+3}^{(-)\ast}& c_{4n+3}^{(+)}c_{4n+2}^{(+)\ast}& \vert c_{4n+3}^{(+)}\vert^2 & c_{4n+3}^{(+)}c_{4n+2}^{(-)\ast} \\
c_{4n}^{(+)}c_{4n+1}^{(+)\ast} & c_{4n}^{(+)}c_{4n}^{(-)\ast} & c_{4n}^{(+)}c_{4n+1}^{(-)\ast} & \vert c_{4n}^{(+)}\vert^2
\end{array}  \right) + \nonumber \\
&& + \left( \begin{array}{cccc}
\vert c_{4n+3}^{(-)}\vert^2 & c_{4n+3}^{(-)}c_{4n+2}^{(+)\ast}  &  c_{4n+3}^{(-)}c_{4n+3}^{(+)\ast} &  c_{4n+3}^{(-)}c_{4n+2}^{(-)\ast}\\
c_{4n}^{(-)}c_{4n+1}^{(+)\ast} & \vert c_{4n}^{(-)}\vert^2 &c_{4n}^{(-)}c_{4n+1}^{(-)\ast}&c_{4n}^{(-)}c_{4n}^{(+)\ast} \\
c_{4n+1}^{(-)}c_{4n+1}^{(+)\ast}& c_{4n+1}^{(-)}c_{4n}^{(-)\ast}& \vert c_{4n+1}^{(-)}\vert^2 & c_{4n+1}^{(-)}c_{4n}^{(+)\ast} \\
c_{4n+2}^{(-)}c_{4n+3}^{(-)\ast} & c_{4n+2}^{(-)}c_{4n+2}^{(+)\ast} & c_{4n+2}^{(-)}c_{4n+3}^{(+)\ast} & \vert c_{4n+2}^{(-)}\vert^2
\end{array}  \right).
\end{eqnarray}
Figure \ref{fig:Fig2} and Fig. \ref{fig:Fig3}  presents the dynamics in the USC and DSC regime.
In the USC regime it is important to note that the oscillations in the mean photon number, Fig. \ref{fig:Fig2}(a), and the population inversion, Fig. \ref{fig:Fig2}(b), point to a decomposition in the normal mode basis involving a high number of components. 
While the strength of the couplings induces high values of von Neumann entropy, Fig. \ref{fig:Fig2}(c), and bipartite concurrence, Fig. \ref{fig:Fig2}(d). 
In the DSC regime, the well defined oscillations in the mean photon number, Fig. \ref{fig:Fig3}(a), point to a localized decomposition in normal modes. 
The same behavior is seen at a lower time-scale in the population inversion,  Fig. \ref{fig:Fig3}(b), which presents a series of oscillations that remind of collapse-revivals until it localizes in the long time. 
The quantum correlations between the qubit-field and qubit-qubit increase and build up in a shorter time with respect to the values typically found in the weak-coupling regime but decrease with respect to the USC regime,  Figs. \ref{fig:Fig3}(c) and \ref{fig:Fig3}(d). 
The increase in the quantum correlations between the field and the qubit ensemble is expected from the increase in the strength of the interaction between them but the peculiar decreasing of the qubit-qubit correlations is not expected because individual members of an ensemble of identical qubits become entangled with one another thanks to the interaction with the field mode, cf. \cite{Tessier2003p062316} and references within.
Our results point to the fact that qubit-qubit entanglement for an initial separable state diminish and become highly localized in time with an increasing coupling of the field. 
This may affect the way we design two-qubit gates in the different regions of the strong coupling regime.

%
\section{Conclusion}

Our main results are the following. 
It is possible to obtain accurate eigenvalues from the RWA for one coupling parameter value in the USC regime if the other coupling parameter value is well into the weak-coupling regime or for both coupling parameters by considering symmetrical detuning for the qubits.
In the DSC regime we have calculated the eigenvalues up to second-order perturbation correction.
Well into the DSC regime, the eigenvalues of the two-qubit Rabi model are described by two forced oscillator spectral series; this is similar to what happens in the single-qubit case \cite{Schweber1967p205, Tur2001p899}.
In the intermediate regime, partitioning the Hilbert space in even and odd parity subspaces simplifies the numerical calculation of the eigenvalues.
The numerical spectra point to energy crossings within and between each of the parity subspaces in the USC regime; the case studied in \cite{Peng2012p365302} and the case of identical qubits seem to be exceptions to this characteristic.
The crossings within the spectra of parity subspaces represent different behavior to that of the single-qubit results, where the spectra branches within each of the parity subspaces do not cross and the parity subspace spectra cross each other \cite{ Tur2001p899, Braak2011p100401}.
During the reviewing period, it has been pointed out that such crossings may appear for the Dicke model describing three identical qubits \cite{Braak2013}.
We have obtained the normal modes for the two-qubit Rabi model by linear algebra methods via parity subspaces as four-term recurrence relations for the coefficients of the eigenstates; in the case of identical couplings, the linear system becomes singular and we cannot provide an analytic solution; nevertheless, it is possible to obtain numeric solutions for this particular case. 
In the Bargmann representation, we find that the recurrence relations for the continued proper functions coefficients are of five terms and reduce to three-term recurrence relations in the case of identical qubits; additionally, for identical qubits, one of the proper functions has a well-defined parity.
We also extend an algebraic approach to find the evolution operator of the system in the RWA. 
Finally, we wanted to show how powerful the parity decomposition is by calculating the time evolution of quantities related to the degree of entanglement in the ensemble-field and qubit system for a simple initial  state of the qubits in their ground state and the field in a coherent state with two photons on average in the USC and DSC regimes, respectively.
We show that while qubit ensemble-field quantum correlations increase with an increasing coupling parameter, qubit-qubit entanglement measured by the concurrence diminishes and localizes in regions of the time evolution.

%

%
\section*{References}
\providecommand{\newblock}{}

\end{document}